\documentclass[times]{TRR}

\usepackage{moreverb,url}

\usepackage[colorlinks,bookmarksopen,bookmarksnumbered,citecolor=red,urlcolor=red]{hyperref}

\usepackage{graphicx}
\graphicspath{ {.}}
\usepackage{caption}
\usepackage{subcaption}
\usepackage{xcolor}
\usepackage[normalem]{ulem}

\newcommand\BibTeX{{\rmfamily B\kern-.05em \textsc{i\kern-.025em b}\kern-.08em
T\kern-.1667em\lower.7ex\hbox{E}\kern-.125emX}}

\begin{document}

\runninghead{Pierce, Hu, Shafi, Boral, Anisimov, Nevo, and Chen}

\title{Accelerating Physics Simulations with TPUs: An Inundation Modeling Example}
\bigskip
\author{Damien Pierce\affilnum{1}\affilnum{2}, R. Lily Hu\affilnum{1}\affilnum{2}, Yusef Shafi\affilnum{1}, Anudhyan Boral\affilnum{1}, Vladimir Anisimov\affilnum{1}, Sella Nevo\affilnum{1}, and Yi-fan Chen\affilnum{1}}
\bigskip

\affiliation{\affilnum{1}Google Research\\
\affilnum{2}Both authors contributed equally to this research}

\corrauth{Damien Pierce, dmpierce@google.com\\
Lily Hu, rlhu@google.com\\
}

\begin{abstract}

Recent advancements in hardware accelerators such as Tensor Processing Units (TPUs) speed up computation time relative to Central Processing Units (CPUs) not only for machine learning but, as demonstrated here, also for scientific modeling and computer simulations. To study TPU hardware for distributed scientific computing, we solve partial differential equations (PDEs) for the physics simulation of fluids to model riverine floods. We demonstrate that TPUs achieve a two orders of magnitude speedup over CPUs. Running physics simulations on TPUs is publicly accessible via the Google Cloud Platform, and we release a Python interactive notebook version of the simulation.
\end{abstract}

\maketitle

\section{Introduction}

The computational needs of deep learning have driven the pursuit of faster hardware accelerators with larger memory. This has led to the development of Tensor Processing Units (TPUs). While TPUs were specifically designed to accelerate machine learning computations, they are sufficiently general to be used for other computing tasks, including scientific computing such as simulations. In this paper we explore and quantify the performance of TPUs for physics simulations. To elucidate TPUs' capabilities quantitatively, we demonstrate their application to riverine flood simulations, which can be used to forecast floods.

To investigate the suitability of TPUs for physics simulation, we implement and run simulations on CPUs and TPUs at various grid resolutions. Resolutions, and thus simulation size, are varied by sampling a digital elevation model of the terrain of the region under study. We first demonstrate the significant speed up obtained using a single TPU core compared to a CPU core for equivalent physical simulations. Then, we examine how the performance of simulations on TPUs scales with an increasing numbers of cores.

We highlight the application of simulating riverine floods owing to the fact that floods are among the most common and most deadly natural disasters in the world, affecting hundreds of millions of people, causing thousands of fatalities annually, and resulting in tens of billions of dollars worth of damages \cite{who2014} \cite{naturaldiasters2017} \cite{UNISDR2015} \cite{jonkman2005global}. Furthermore, the statistics have continued to increase in recent decades \cite{kundzewicz2014flood}. The United Nations noted floods to be one of the key motivators for formulating the Sustainable Development Goals (SDGs), challenging us that "(we know) that earthquakes and floods (are) inevitable, but that the high death tolls (are) not" \cite{UNSDG}. Flood forecasting early warning systems are an effective mitigation tool, reducing fatalities and economic costs by about a third and in some cases almost a half \cite{who2014} \cite{worldbank2011}\cite{pilon2002guidelines}.

\begin{figure}
  \centering
  \includegraphics[scale=0.135]{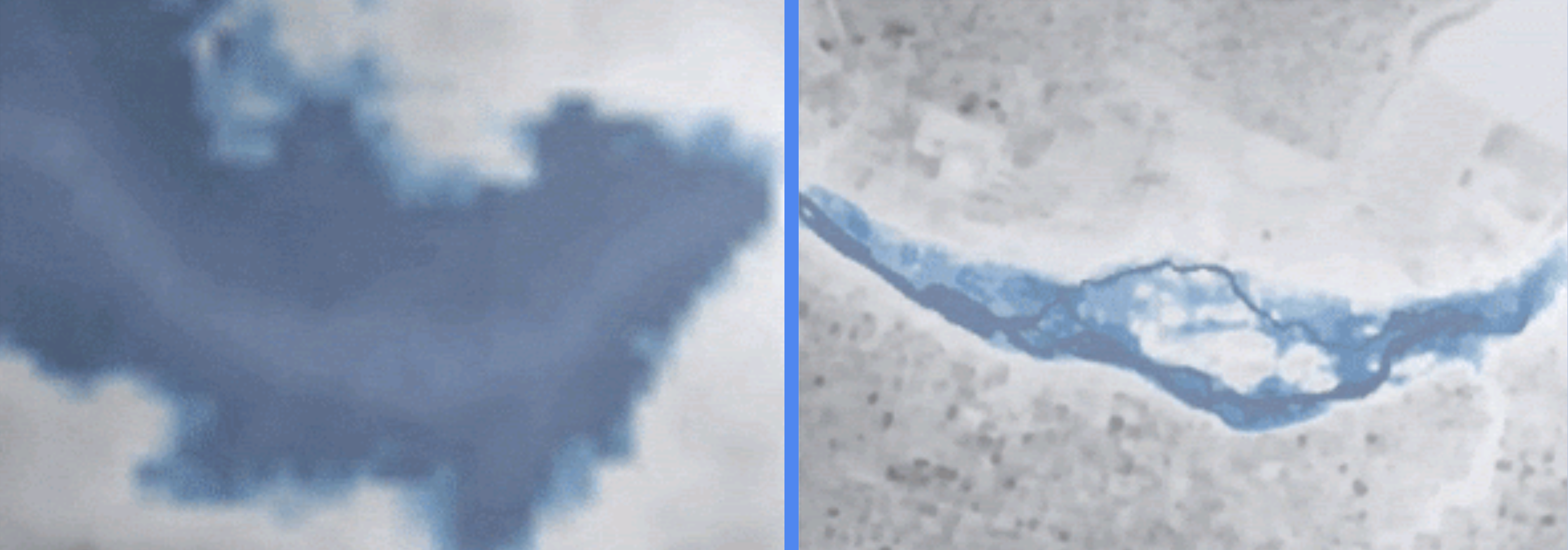}
  \caption{A motivating example where a flood prediction at low spatial resolution (left) differs from the prediction at high spatial resolution (right).}
  \label{fig:low_vs_high_res}
\end{figure}

Higher spatial accuracy is vital both in allowing disaster management agencies to plan effective mitigation and relief efforts, and in providing actionable information to individuals. Higher resolutions aid in incorporating nuances, e.g., embankments, river bathymetry, etc. Different spatial resolutions can lead to significant differences in the accuracy of the resulting water height. A motivating example of this difference is shown in Figure \ref{fig:low_vs_high_res}. Increasing spatial resolution increases computational resource requirements, thereby potentially becoming prohibitively expensive and/or reducing the ability to respond to flood events in resource-constrained environments where large numbers of CPUs may be unavailable. However, TPUs publicly available on Google Cloud Platform offer a flexible and accessible mechanism for harnessing distributed computation at high resolution while minimizing the need for specialized high performance computing (HPC) hardware.

The rest of the paper is organized as follows. The Background section provides background on TPUs themselves and riverine flood forecasting. The implementation of physics simulations on TPUs is described in the Methods section. The Discussion section presents and analyzes experiments comparing TPUs with CPUs as well as scaling properties of TPUs, and discusses further use cases. We also publicly release code\footnote{\url {https://github.com/google-research/google-research/blob/master/simulation_research/flood/}} to run the simulations on TPUs, which are available through the Google Cloud Platform.

\section{Background}\label{background}

\subsection{Tensor Processing Units}

A new class of hardware accelerators recently emerged, primarily targeting computations for large-scale deep learning. These new hardware accelerators include the Tensor Processing Unit (TPU) \cite{Jouppi2017} developed by Google. A single TPU v3 offers $420 \times 10^{12}$ floating-point operations per second (FLOPS) and 128 GB of high-bandwidth memory (HBM). Furthermore, multiple (2048) TPU cores can be connected through a high-speed toroidal mesh network into a ``pod'', achieving 100+ peta-FLOPS and 32TB of HBM.

Despite having originally been designed for deep learning, TPUs have successfully been applied outside of machine learning to numerous scientific computation tasks conventionally requiring specialized high-performance computing resources. These include Monte Carlo simulation of Ising models \cite{tpu_ising_sc19}, MRI reconstruction \cite{simresearchmri}, Discrete Fourier Transform \cite{Lu2020LargeScaleDF}, and nonuniform Fast Fourier Transform \cite{ma2021fft}. A common property in these applications is their use of large, high-resolution grids: TPU pods constitute a class of interconnected accelerators particularly well-suited for such settings. TPUs have also been shown to be viable for financial Monte Carlo\cite{belletti2020tensor} problems including parallelized pseudo-random number generation and options pricing.

The programming model for TPUs typically makes use of high-level Python code using TensorFlow \cite{abadi2016tensorflow} or JAX \cite{jax2018github}. Such high-level APIs interface with the lower level Accelerated Linear Algebra (XLA) compiler that converts the higher-level operations to an optimized form readily executed on TPU cores. This compilation step happens once per program execution, with the low level optimized (LLO) code executed repeatedly or iteratively as specified by the Python code.

One of the key advantages of using TPUs is the high bandwidth inter-chip interconnect (ICI) which can share information between cores efficiently (4x656 Gbps / chip for Google Cloud TPU v3)\footnote{\url{https://cloud.google.com/tpu/docs/system-architecture}}. TPUs also have reasonable high performance compute (HPC) characteristics, as will be shown in Section Results. Solving PDEs involves large matrix multiplication operations, a capability for which TPUs are especially well-suited owing to their Matrix Unit (MXU) and Vector Processing Unit (VPU).

In our study, we designed and implemented a TensorFlow PDE solver and library to facilitate programming distributed algorithms on TPUs in easy-to-express Python code. The finite difference computation is distributed across multiple cores following the ``single instruction, multiple data'' (SIMD) paradigm of domain decomposition \cite{smith1997domain}. Array slicing operations are used for computing spatial finite differences. The current implementation achieves over 100x speed-up using a single TPU core relative to our reference single CPU core implementation.

\subsection{Riverine Flood Forecasting Methods}

At a high level, a riverine flood forecasting system receives measurements and/or forecasts of the river water levels (i.e. heights) or discharge rates, and predicts the extent of the resulting flooding. These systems usually consist of a few components. First, water level measurements can be collected from stream gauges. Alternatively, a hydrologic model can provide forecasts of river discharges. Another component is an inundation model that takes heights or discharges as input, and provides forecasts of flood extent, water depths and velocities, etc.

There are a wide range of models used for inundation modeling \cite{teng2017flood}. The standard, most researched and commonly used methods are based on hydraulic models, as these types of models tend to be the most accurate and provide full hydrodynamic information. Nevertheless, due to the heavy computational cost of the hydraulic models and their calibration, simplified models are often chosen for inundation modeling \cite{zhang2018comparative, afshari2018comparison}, compromising accuracy and availability of some information.

This paper focuses on an efficient implementation of a fully-2D hydraulic model on TPUs, aspiring to bridge the computational gap and enable more organizations to rely on these models. This is in contrast to combined 1D-2D models which are often used to reduce computational complexity due to computation constraints. We also hope that this work will enable new avenues of research on calibration and applications of hydraulic models.

\section{Methods} \label{method}

Direct numerical integration of the PDE is used to advance the simulation forward in time. To accelerate performance, the computation is parallelized across multiple TPU cores. In the rest of this section, we describe implementation details for our TPU PDE solver. We then outline the method of application of our solver to high resolution riverine simulations, including the acquisition of necessary data for parameterizing the region we study.

\subsection{TPU Implementation}

\begin{figure*}[ht]
  \centering
  \includegraphics[scale=0.4]{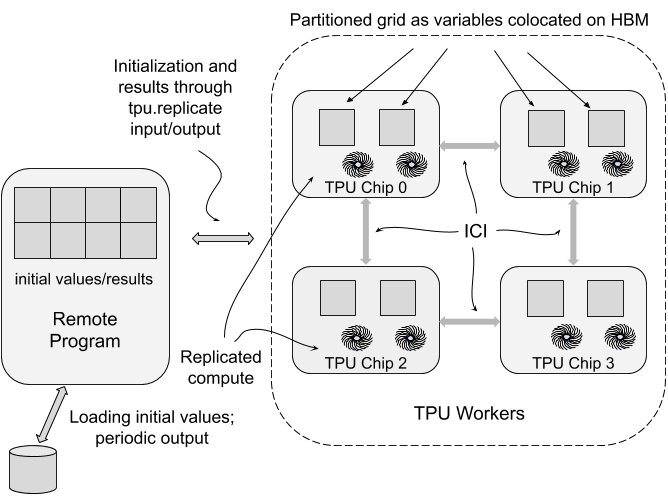}
  \caption{Distribution and parallelization on TPUs}
  \label{fig:distribution}
\end{figure*}

We accelerate the solution of PDEs by distributing the work over multiple TPU cores in a usual SIMD approach. The area of interest is represented on a uniform rectangular grid corresponding to the number of points in the digital elevation model. The grid is divided into subgrids, one for each TPU core (padding is added to the grid so that it can divide evenly into the number of TPU cores). This is illustrated in Fig. (\ref{fig:distribution}), which indicates that \texttt{tpu.replicate} (a TensorFlow operator) is used to send TensorFlow code to the TPU workers. The data corresponding to the state variables are similarly distributed to the subgrids.

The update proceeds as follows. First, for the dynamic flux fields, halo exchange of the boundaries is performed, wherein the 1-pixel-width borders of each subgrid are sent to and received from the nearest neighbors. The TPU API in TensorFlow includes an efficient operation for the exchange of information between cores, CollectivePermute\footnote{\url{https://www.tensorflow.org/xla/operation\textunderscore semantics}}. The communication in halo exchange is facilitated by high bandwidth inter-chip interconnect (ICI). Next, dynamic updates are performed on the dynamic state fields (equations (\ref{eq:st-venant-discrete}) and (\ref{eq:continuity-eq-discrete}) in our application). Finally, halo exchange is done for the height fields $h$. In this case the halo exchange operation includes applying the inflow and outflow boundary conditions.

\subsection{Shallow Water Equations}

Hydraulic models simulate the flow of water in a given topography. When properly calibrated, hydraulic models are capable of accurately predicting inundation extent and water depth \cite{bates2010simple}. The classical approach is to time-integrate the Saint-Venant shallow water PDEs. Ignoring the negligible advection term, the PDEs are given by \cite{vreugdenhil94, de2012improving}
\begin{equation} \label{eq:st-venant_1}
\frac{\partial h}{\partial t} + \frac{\partial q_1}{\partial x} + \frac{\partial q_2}{\partial y} = 0
\quad \end{equation} and
\begin{equation} \label{eq:st-venant_2}
\quad
\frac{\partial q_i}{\partial t} + g h \frac{\partial (h+z)}{\partial x_i} +
\frac{g n^2}{h^{7/3}} \|\mathbf{q}\| q_i = 0, \quad i \in \{ 1, 2 \}
\end{equation}
where $\mathbf{q} = (q_1, q_2) = (q_x, q_y)$ is the flux in the $x$ and $y$ directions, $h$ is the water height, $z$ is the surface elevation (topography), $g$ is the gravitational acceleration constant, and $n$ is the Manning friction coefficient. Equation \ref{eq:st-venant_1} represents conservation of mass: when water flows into a cell, the water height rises. Equation \ref{eq:st-venant_2} represents Newton's second law: water accelerates (its flux increases) due to gravity when there are differences in water height, and decelerates due to friction.

Our implementation is based on the numerical scheme of \cite{de2012improving} for solving the inertial form of the Saint-Venant equations, a canonical approach used in inundation modeling and by the LisFlood-FP system \cite{bates2000simple,bates2010simple}. The region of interest is represented as a staggered 2D grid with square cells, such that there is a water depth node located in the center of each cell, and a flux node located on a link between every two cells. The flux is integrated in time, independently at each axis, following a 1D finite difference discretization of Equation \ref{eq:st-venant_2}, as developed in \cite{de2012improving}:

\begin{equation} \label{eq:st-venant-discrete}
q_i^{t + \Delta t} = \frac{q_i^t - g h_f \frac{\partial{(h+z)}}{\partial{x}_i}}{1 + g\, \Delta t\, n^2 ||\textbf{q}^t|| h_f^{-7/3}}
\end{equation}

where $h_f$ is the difference between the maximum water height and the maximum surface elevation of the adjacent cell, or in other words, the section through which the water can flow. After each time integration of the flux, we apply the conservation of mass equation as described by Equation(\ref{eq:st-venant_1}) to update the water depth in each cell:

\begin{equation} \label{eq:continuity-eq-discrete}
h^{t + \Delta t} = h^t + \Delta t \ \frac{(q_x^- - q_x^+) + (q_y^- - q_y^+)}{\Delta x}
\end{equation}
where $q_x^-(x_0, y_0)$ corresponds to the flux node along the x-axis from the previous cell to cell $(x_0, y_0)$. Similarly, $q_x^+(x_0, y_0)$, $q_y^-(x_0, y_0)$, $q_y^+(x_0, y_0)$ are the flows to the next cell along the x-axis, flows from the previous cell along the y-axis, and flows to the next cell along the y-axis to cell $(x_0, y_0)$, respectively. While we also implemented variable time step treatments that can improve stability and/or execution time, for simplicity's sake we use a constant time step in the results shown here.

Note that $q_x^-(0, y_0)$ cannot be directly computed by (\ref{eq:st-venant-discrete}), as it relies upon the left adjacent water height cell, which doesn't exist. This is true for any boundary flux cell, and hence these values must be computed separately through boundary conditions. We use two types of boundary conditions for $h$: the inflow boundary receives a predetermined amount of flux, and the outflow boundary conditions are computed using Manning's equation \cite{akan2011open}.

To describe the boundary conditions in detail, we derive a local approximation of Manning's equation, analogously to the numerical scheme of the Saint-Venant equation \cite{de2012improving}, by approximating the hydraulic radius $R$ with the water depth of a cell. This leads to the following equation for a single grid cell:

\begin{equation} \label{eq:mannings-eq-discrete}
Q = \frac{A h^{\frac{2}{3}} \sqrt{S}}{n}
\end{equation}
where $Q$ is the flux and $S$ is the slope of the channel and $A$ is the cell area ($=(\Delta x)^2$ in case of square cells).

The inflow boundary conditions are specified by three parameters:
\begin{itemize}
    \item A continuous cross section through which the inflow enters the computation grid ($C_{in}$). The cross section will typically cover the channel and some margin to account for increasing channel width as the flux increases.
    \item The flux through the given cross section at each point in time of the simulation ($Q_{in}$). The flux can be a constant function for steady-state simulations, or a measured/predicted hydrograph in case it is available.
    \item The slope of the channel to use in Manning's equation ($S_{in}$). The slope is usually inferred from the average channel slope near the inflow boundary condition.
\end{itemize}
At each time integration step the input flux is distributed along the cross section cells by solving the inverse Manning's equation (\ref{eq:mannings-eq-discrete}). In other words, we find a water level $W$ s.t.

\begin{equation} \label{eq:inflow-boundary-condition}
\sum\limits_{c \in C_{in}}q_{in}(c) = Q_{in}
\end{equation}

\begin{center}
where  $q_{in}(c) = \frac{A \hat h(c)^{\frac{2}{3}} \sqrt{S_{in}}}{n}$ and $\hat h(c) = max(W - z(c), 0)$.
\end{center}

The outflow boundary conditions are specified by two parameters:
\begin{itemize}
    \item A continuous cross section through which the water is allowed to flow out ($C_{out})$. As in the inflow boundary, this cross section will typically cover the channel and some margin around it.
    \item The slope of the channel to use in Manning's equation ($S_{out}$). As in the inflow boundary, the slope is usually inferred from the average channel slope near the outflow boundary condition.
\end{itemize}
The output flux is computed separately at each output cell according to Manning's equation (\ref{eq:mannings-eq-discrete}). To create the simulation, the above equations are coded in Python using TensorFlow.

\subsection{Data}

\begin{figure}[t]
    \centering
    \includegraphics[width=.405\textwidth]{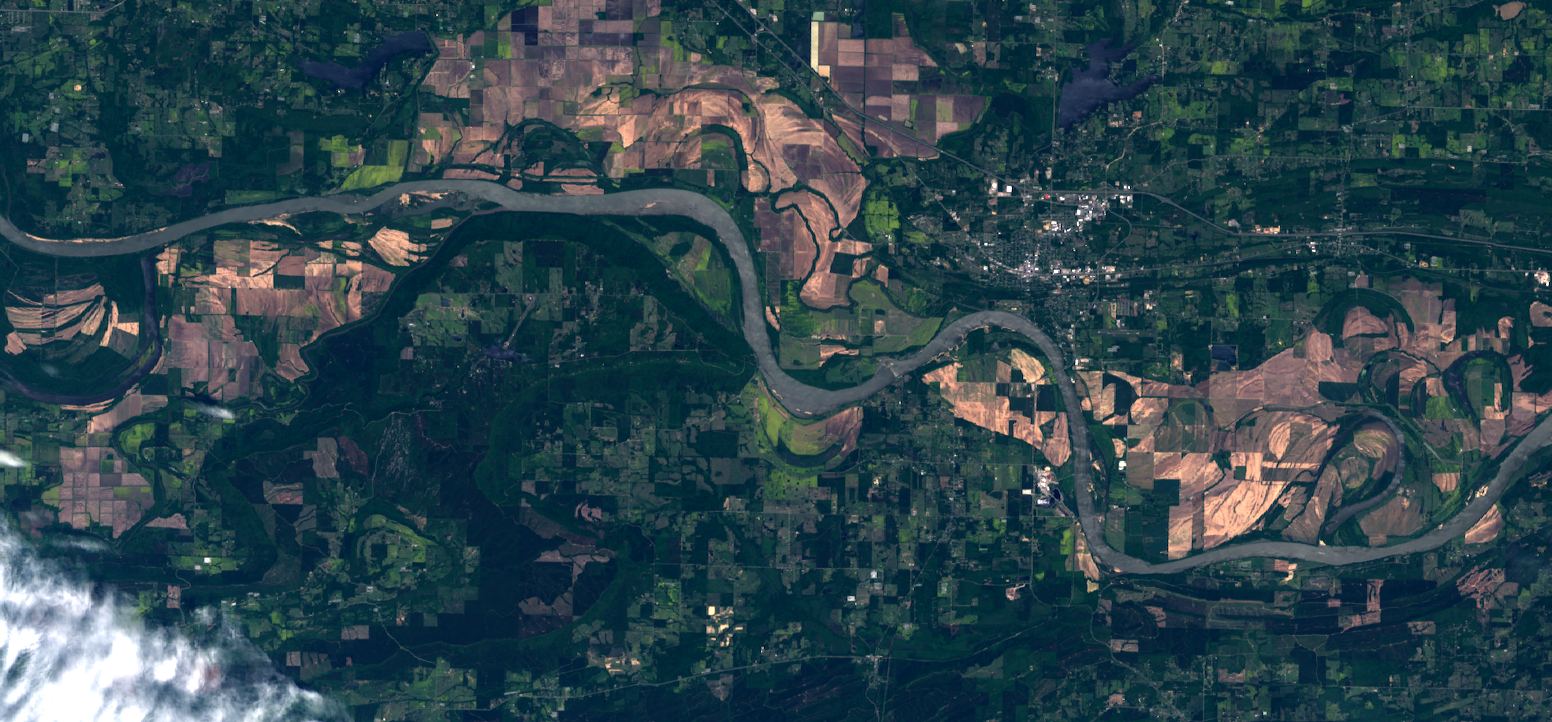} 
    \caption{Satellite imagery of the simulation area corresponding to a section of the Arkansas River during non-flood conditions. The white section in the bottom left of the image is clouds.}
    \label{fig:nonflood-imagery}
\end{figure}

For our flood simulation experiments, we selected a 70 km stretch of the Arkansas River, located between the Arkansas cities Fort Smith and Conway. This area was severely flooded during parts of May and June 2019. The corresponding satellite imagery for this simulation area is shown in Figure \ref{fig:nonflood-imagery}, taken a few days before the flood that we simulated. During this flood event, aerial imagery was collected by the Arkansas Department of Transportation and is available online on the gis.arkansas.gov platform. We compare this imagery with the simulated flood extent in Section 4.1.

\begin{figure}[t]
    \centering
    \includegraphics[scale=.25]{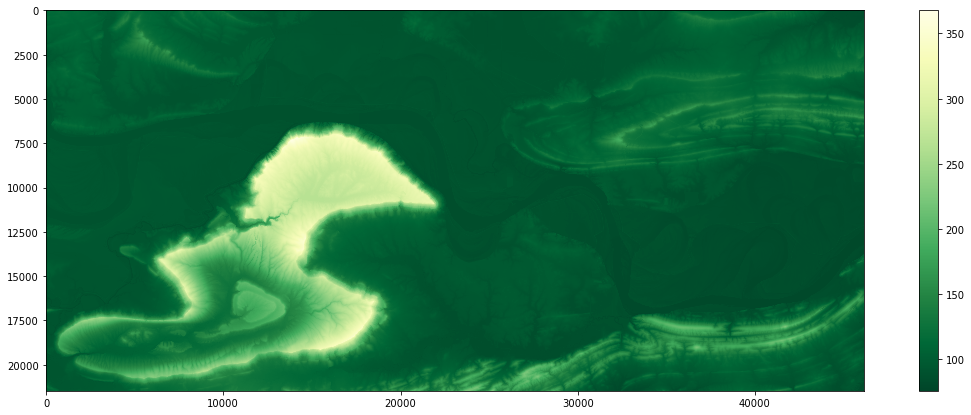} 
    \caption{Digital elevation map at 1 meter resolution of the simulation area corresponding to a section of the Arkansas River. The colors correspond to elevation in meters.}
    \label{fig:dem}
\end{figure}

Probably the most critical component for hydraulic modeling is an accurate digital elevation model (DEM). In this paper we use a DEM that was downloaded from the publicly available USGS 3D Elevation Program (3DEP) \cite{usgs20173dep}. For the chosen area the 3DEP dataset provides a LiDAR DEM at 1 meter resolution. The raw data was collected during the years of 2014-2016. The DEM is visualized in Figure \ref{fig:dem} and covers an area of 21471 meters by 46129 meters.

This level of detail plays an important role in the ability of the model to accurately simulate floods \cite{meesuk2015urban}, but comes with an expensive computation cost, as the run-time complexity of a two-dimensional hydraulic model is inversely proportional to the cube of the resolution. We run simulations using 1, 2, 4, and 8 meter DEMs. The coarser DEMs were created from the 1 meter DEM via averaging.

\subsection{Running the Simulation}

When simulating a large river flood event as we do here, it is a good approximation to run the simulation long enough so that it approximately reaches steady state. This is achieved after running the simulation for two "simulation days". We use a $\Delta t$ of 0.1 second, so the results shown correspond to running the simulation for 1.728 million time steps.

The primary parameters of the model include the slopes at the inflow and outflow boundaries, the Manning coefficients, and the discharge at the inflow boundary. The Manning coefficients and slopes were set to canonical values. The remaining parameter that affects the simulation then is the discharge rate at the inflow boundary. This is chosen such that the flood extent in the simulation after running for 2 "simulation days" closely matches the aerial imagery.

\section{Results}\label{discussion}

In this section we show the results of the simulation, and compare the results to aerial imagery to study accuracy qualitatively. We then compare performance of a single CPU vs. TPU core, and next discuss how the TPU results scale with an increasing number of cores. Finally, possible extensions and use cases are described. The calculations in all cases are done using single precision (32 bit) floating point arithmetic. There is currently no support for 64 bit on TPU.

\subsection{Comparison Between Simulation and Flood Aerial Imagery}

\begin{figure*}
 \begin{subfigure}[ht]{1\textwidth}
 \centering
    \includegraphics[width=0.9\textwidth]{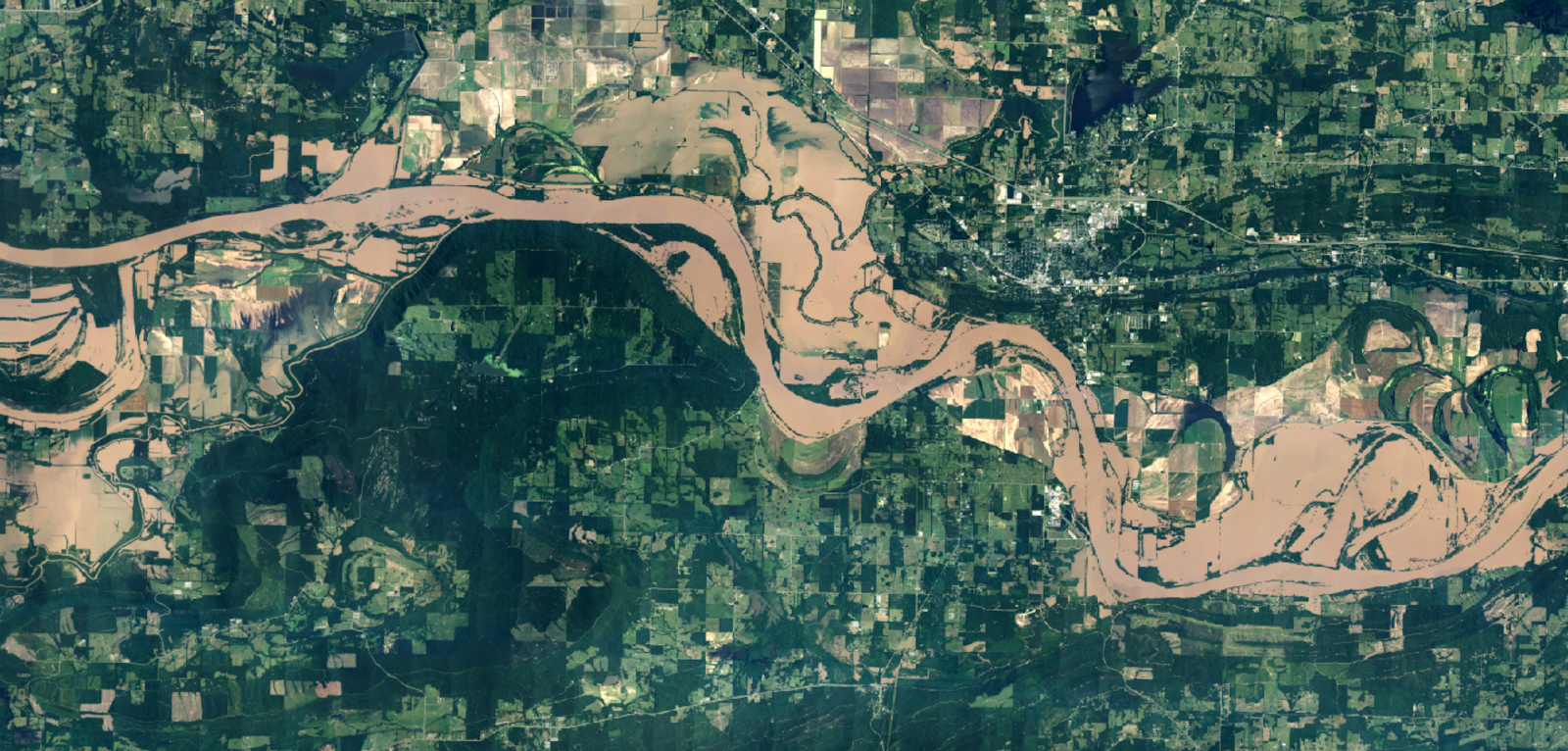}
 \caption{Aerial imagery.}
 \end{subfigure}%

  \vskip\baselineskip

  \begin{subfigure}[ht]{1\textwidth}
  \centering
  \includegraphics[width=0.9\textwidth]{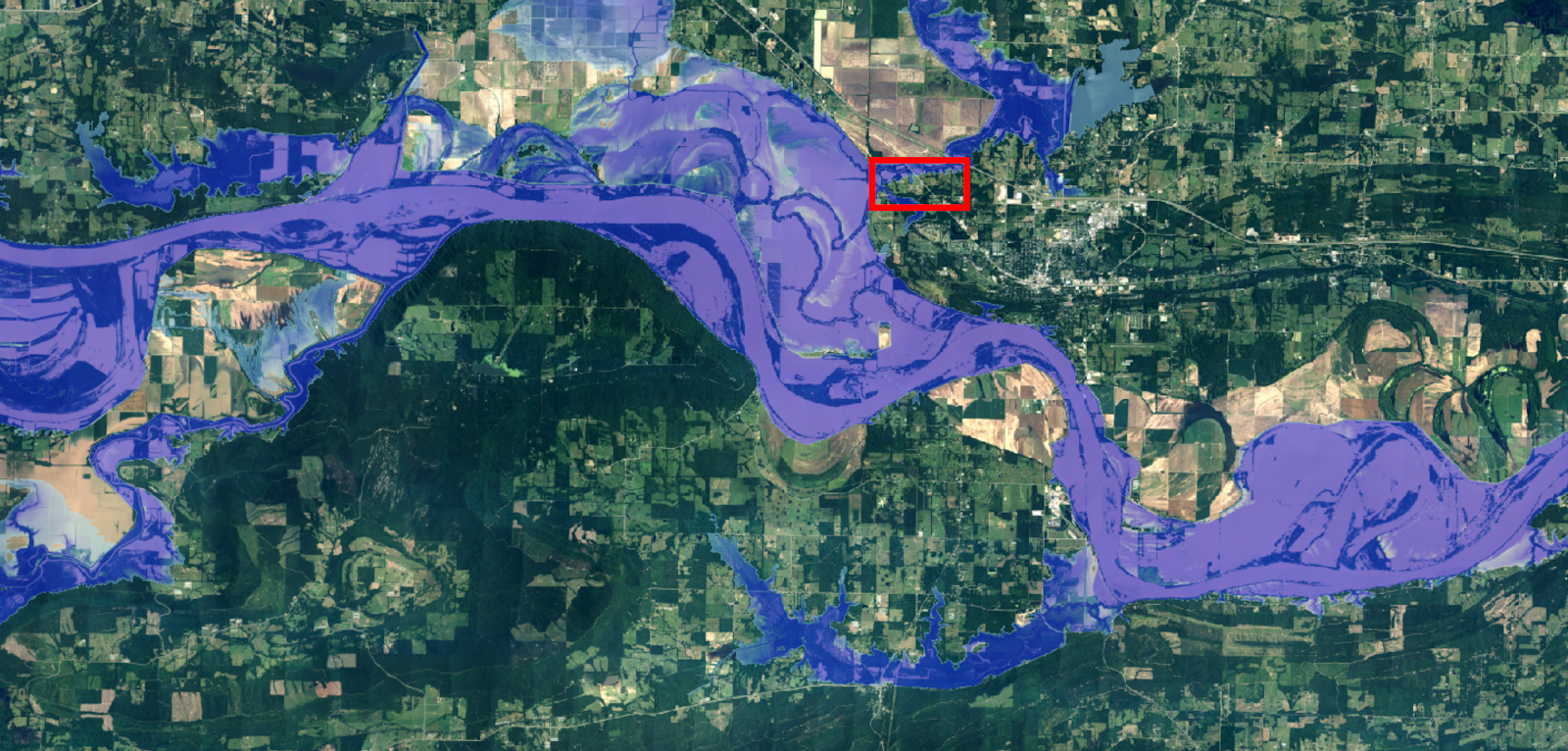}
  \caption{Simulation result.}
  \end{subfigure}%
\caption{Comparison of the aerial imagery (a) versus the simulation result on a 1 m grid superimposed on the aerial imagery (b), with the red rectangle identifying the zoomed in area displayed in the Figure \ref{fig:aerial-zoomedin}.}\label{fig:aerial-zoomedout}
\end{figure*}

As discussed above, after setting other simulation parameters to canonical values, the only parameter left to be determined is the discharge at the inflow boundary. We ran the simulation for two "simulation days" at various discharge rates and compared the flood extents in the simulations with aerial imagery. A discharge rate that matched the imagery well was 15000 $m^3/s$. Later we discovered a USGS reference \cite{usgs} which lists some streamgage measurements during this flood. The streamgage closest to our simulated region gave a peak discharge value of 16000 $m^3/s$, which is quite close to the value we determined in our simulation, especially considering that the day of the peak streamgage measurement (5/30/2019) was not exactly the day we chose for satellite imagery (5/19/2019), the location (Dardanelle) was about 12 km upstream of our simulated region, and the simulation was not otherwise calibrated.

\begin{figure*}
 \begin{subfigure}[b]{1\textwidth}
 \centering
    \includegraphics[width=0.9\textwidth]{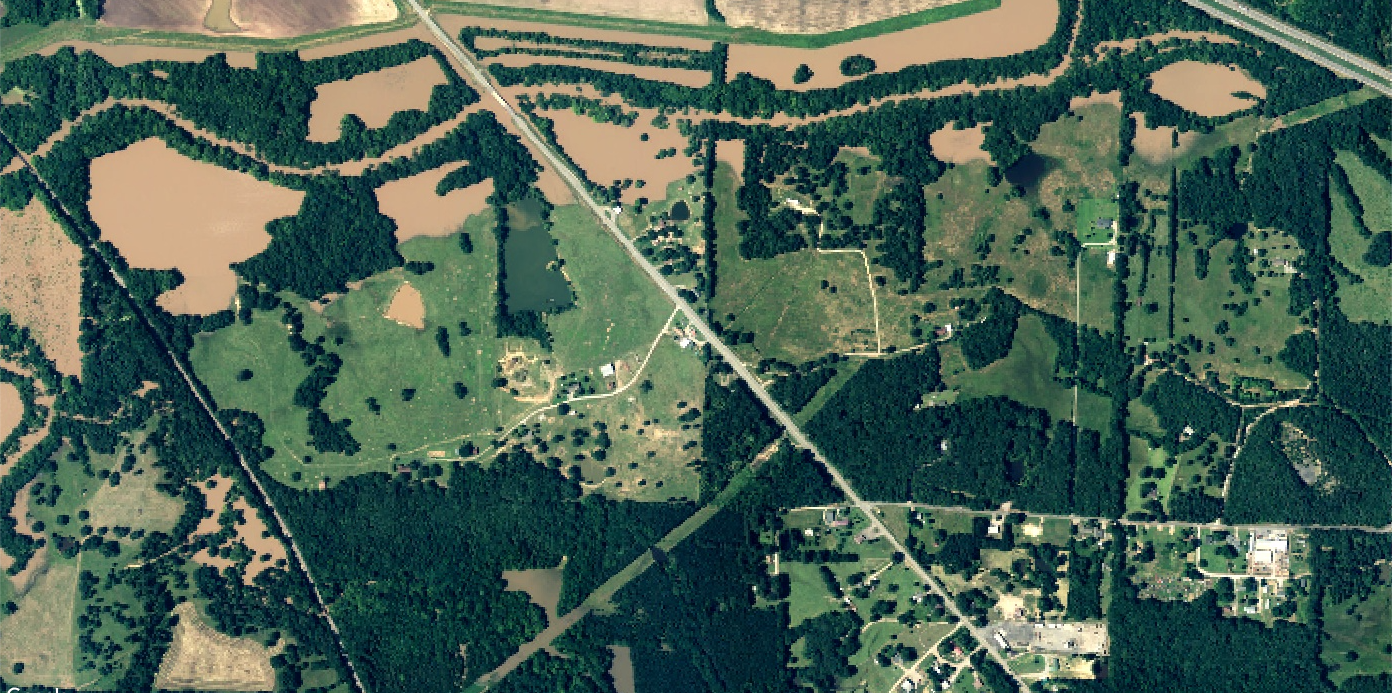}
 \caption{Aerial imagery.}
 \end{subfigure}%

 \vskip\baselineskip

  \begin{subfigure}[b]{1\textwidth}
  \centering
  \includegraphics[width=0.9\textwidth]{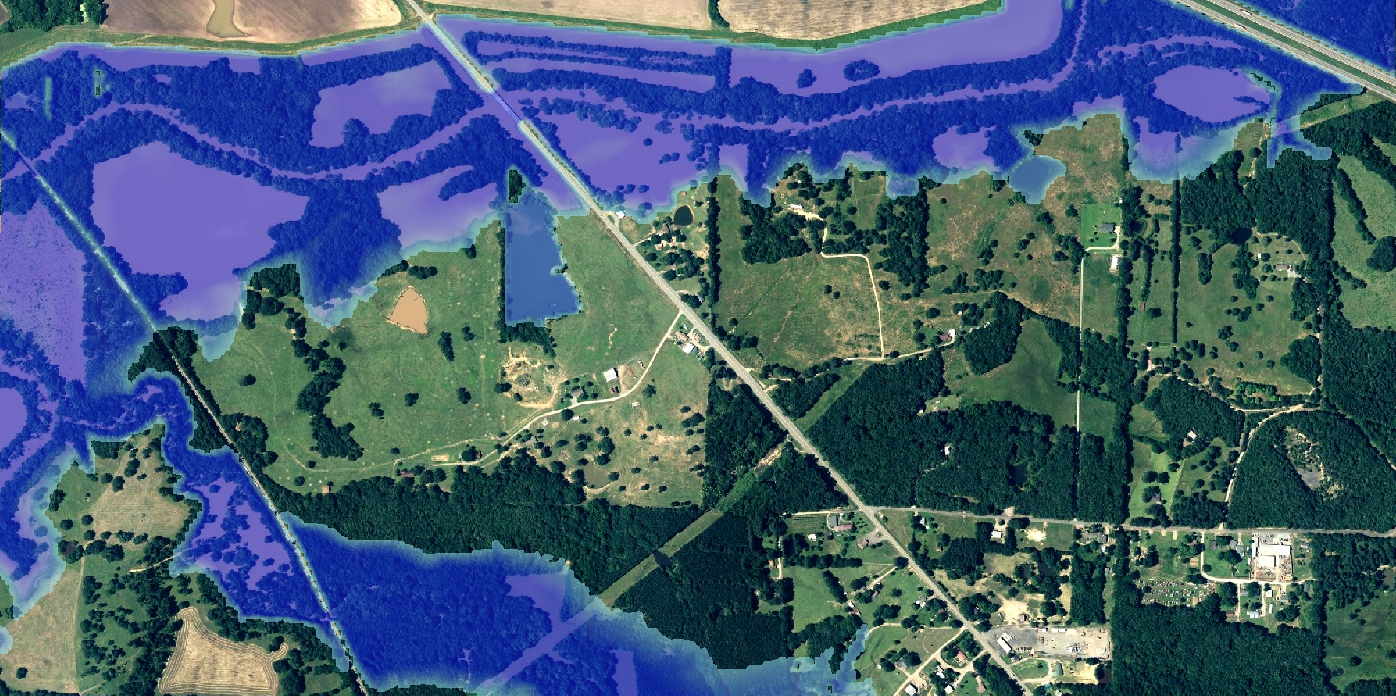}
  \caption{Simulation result.}
  \end{subfigure}%
\caption{Close up comparison of the aerial imagery (a) versus the simulation result on a 1 m grid (b).}
\label{fig:aerial-zoomedin}
\end{figure*}

Figure \ref{fig:aerial-zoomedout}(a) displays aerial imagery of the flood. The flooded areas are brown from the flood water. Some of the adjacent brown areas are dirt just as in the non-flood image in Figure \ref{fig:nonflood-imagery}. Figure \ref{fig:aerial-zoomedout}(b) shows simulation results with a discharge of 15000 $m^3/2$ as a blue mask superimposed on the aerial imagery.  Looking at the entire simulated region, general agreement can be seen between the two figures.

Agreement can be seen in much more detail if we zoom in to the region indicated by the red rectangle. This smaller region is shown in Figure \ref{fig:aerial-zoomedin}, again with the aerial imagery in Figure \ref{fig:aerial-zoomedin}(a) and the simulation result superimposed in Figure \ref{fig:aerial-zoomedin}(b). The simulation agrees well. At first glance, it may appear that the simulation incorrectly flooded into dark green areas. However, upon closer inspection of the aerial imagery, these dark green areas are actually trees that are visually occluding the flood water beneath their canopy. The DEM correctly specifies the ground beneath the tree canopy and hence the simulation appropriately floods some of the tree-covered areas.

\begin{figure}
 \begin{subfigure}[b]{0.48\textwidth}
    \includegraphics[width=1\textwidth]{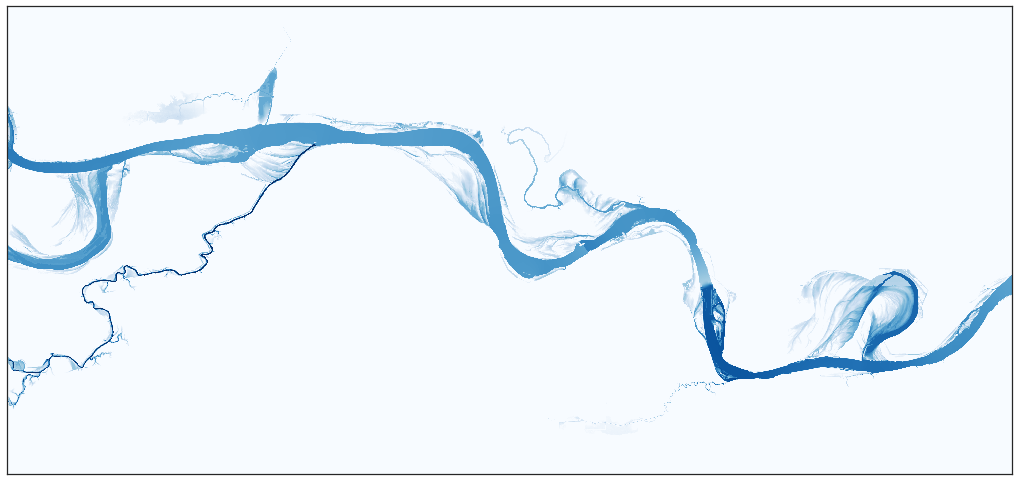}
 \caption{3000 $m^3/s$ discharge}
 \end{subfigure}
\hfill
 \begin{subfigure}[b]{0.48\textwidth}
    \includegraphics[width=1\textwidth]{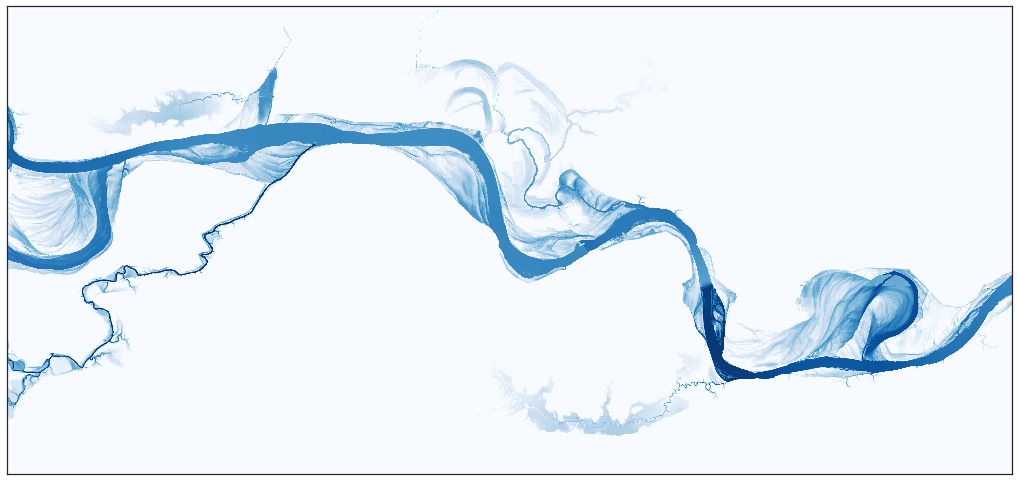}
 \caption{7000 $m^3/s$ discharge}
 \end{subfigure}

\vskip\baselineskip

 \begin{subfigure}[b]{0.48\textwidth}
    \includegraphics[width=1\textwidth]{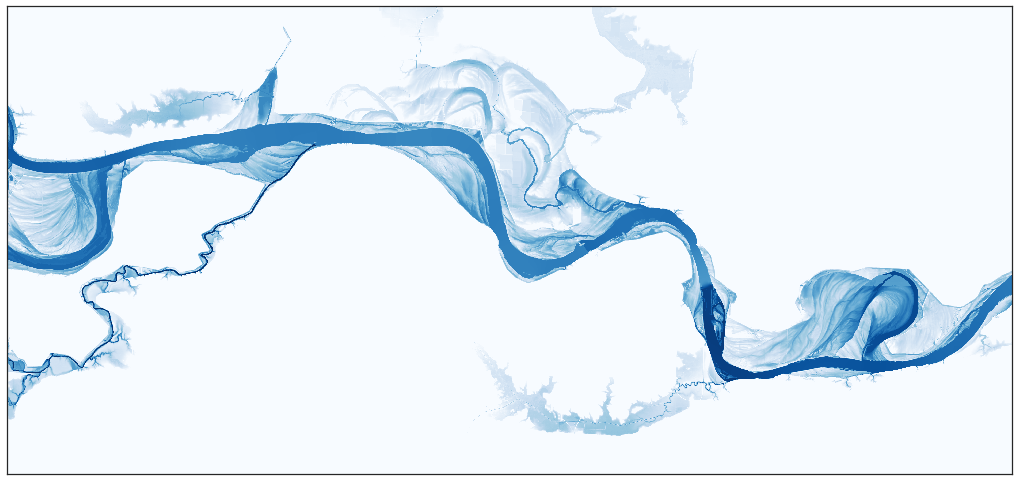}
 \caption{11000 $m^3/s$  discharge}
 \end{subfigure}
\hfill
 \begin{subfigure}[b]{0.49\textwidth}
    \includegraphics[width=1\textwidth]{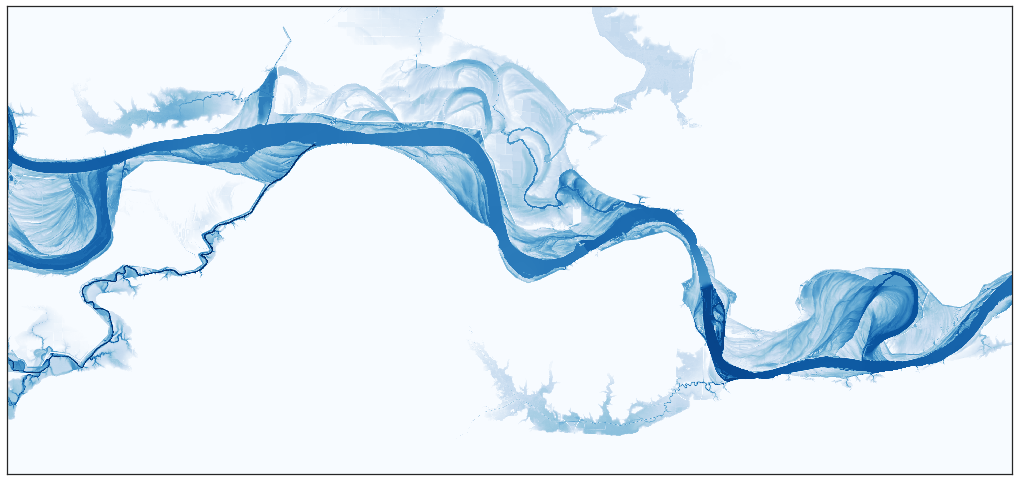}
 \caption{15000 $m^3/s$ discharge}
 \end{subfigure}

\caption{Simulated Arkansas River water extent images for four discharges on a 1 m grid. (a) shows the extent for a normal discharge, (d) for the May/June 2019 flood discharge. (c) and (d) show the extents for intermediate discharges.}
\label{fig:extent}
\end{figure}

The progression of flooding is shown in Fig. \ref{fig:extent}. We show the water extent after running the simulation until it approximately reaches steady state (2 simulation days) for four discharges. The upper left image is the result for a typical amount of discharge, 3000 $m^3/s$. The result corresponding to the May 2019 flood (15000 $m^3/s$) is shown in the lower right image. The other two show the results for the intermediate discharges.

\subsection{Comparison of CPU and TPU}

\begin{table*}
\small\sf\centering
\caption{CPU-TPU Comparison Results Per Core}
\label{table:CPU-TPU}
 \begin{tabular}{ccccc}
 \toprule
 Cell Size (m) & Grid Points (M) & CPU steps/s & TPU steps/s & Speed Up \\ [0.5ex]
 \midrule
 2 & 248 & 0.014 & 7.57 & 540 \\
 4 & 62 & 0.05 & 30.2 & 604 \\
 8 & 15 & 0.21 & 118 & 562 \\
 \bottomrule
\end{tabular}
\end{table*}

For the CPU / TPU comparison we focus on the dynamic time step update calculation, removing communication by running on a single core. Table (\ref{table:CPU-TPU}) shows the number of steps calculated per second achieved on an Intel Xeon(R) Platinum 8273CL 28 Core processor running at 2.2 GHz that can provide about 2 TFLOPS\cite{cpu_flops}, and on the Google Cloud TPU v3. Also calculated in the table is the relative speed up for three resolutions. The CPU code was ported to TPU, replacing code that distributes on CPU with code that distributes to TPU. The CPU implementation uses Eigen \cite{eigenweb} and the TPU implementation uses XLA.

For all resolutions the TPU v3 runs a simulation step over 500 times faster than the CPU. This demonstrates a huge boost in speed when running the simulation on TPU: a simulation that takes one day on a CPU can take less than three minutes on a TPU v3. However, this comparison is somewhat artificial since (whether using CPUs or TPUs) multiple cores are typically used in practice.  In the next section we show how the TPU results scale with an increasing number of cores.

\subsection{Scaling on TPU}

TPUs are organized into connected racks called pods. A TPU v3 pod contains 2048 TPU cores. A simulation can run on a full pod or a subset, called a slice. In the following we show results when running on various slices or fractions of slices, from a single TPU core to one quarter of a pod: 512 cores.

\begin{figure}
    \centering
    \includegraphics[scale=.37]{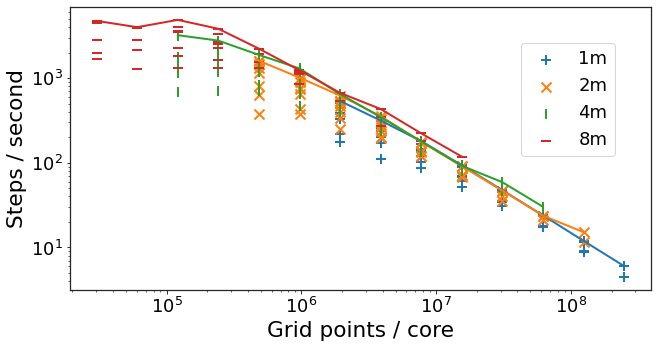}
    \caption{Number of simulation steps per second versus number of grid points per core. The higher the number of steps per second, the better.}
    \label{fig:tpu-scaling}
\end{figure}

We ran the simulation for four resolutions: 1, 2, 4, and 8 meter. The number of grid points varies from 15.5 million (8 meter resolution) to one billion (1 meter resolution). Figure (\ref{fig:tpu-scaling}) shows combined scaling results from all the runs at the various resolutions and various numbers of cores. This is achieved by plotting the number of simulation steps per second vs. the number of grid points per core. The extrema of the x-axis extend from $3 \times 10^4$ grid points per core (corresponding to the 8 meter simulation on 512 cores), to  $2.5 \times 10^8$ (corresponding to the 1 meter simulation running on 4 cores). The best scaling results for each resolution are plotted as lines. These lines overlap to a large extent, indicating high weak scaling efficiencies.

If we fix the number of TPU cores and the resolution, we find a varying number of steps achieved per second. This is because different efficiencies are achieved depending on how a full grid is divided up into subgrids. For example, with 8 cores, one could divide the grid into $2 \times 4$ subgrids or $8 \times 1$ subgrids (among others). The various subgrid choices result in a greater or fewer number of neighbors, and so a greater or fewer number of communications per time step (halo exchanges). The subgrid choices that minimize the number of neighbors (and so minimize the number of communications) are the most extreme layouts: $1 \times N$ or $N \times 1$ for $N$ subgrids (for $N$ TPU cores). These two layouts typically give similar results. The worst subgrid choices correspond to an equal or close to equal number of subgrids per axis (whatever is close to $\sqrt{N} \times \sqrt{N}$), which maximizes the number of communications. The worst subgrid choice can run up to 4.7 times more slowly than the best.

\begin{figure}
    \centering
    \includegraphics[scale=.37]{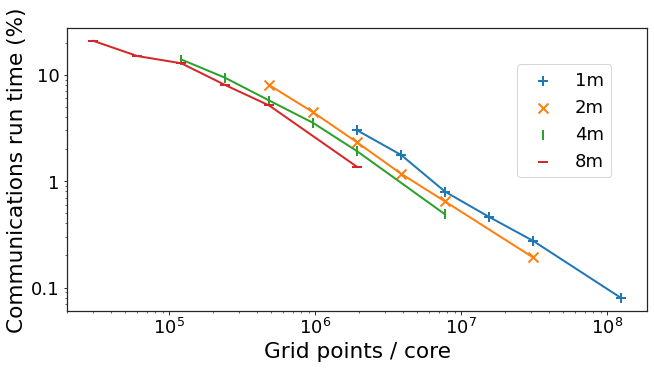}
    \caption{Communication time in percent versus number of grid points per core.}
    \label{fig:comms-percent}
\end{figure}

In high performance computing, increasing the number of cores at a fixed resolution will in general speed up a computation because the number of points per core is reduced. This speed up is offset somewhat by the increase in communication time that comes with more cores. In Figure (\ref{fig:comms-percent}) we show the decrease in communications time as the number of grid points per core is increased (so, the number of cores is decreased) for the various resolutions. The communications time shown is the average time it takes to perform the TPU \texttt{CollectivePermute} operations (which transfer data between the cores) as a percentage of the step time. (The average over various subgrid choices is taken.) The percentage of the step time taken by communication ranges from less than ${1/10}$ of a percent for the highest 1 meter resolution on 8 cores, to over 20\% for the fewest grid points per core (8 meter resolution with 512 cores).

The times it takes to compute 1 million steps for the various resolutions are shown in Table (\ref{table:1M-steps}) for various numbers of cores. These results are for the most efficient subgrid layouts. The coarsest 8 meter simulation can compute one million steps in 6 minutes on 512 cores vs. 43 minutes on 8 cores. The finest 1 meter simulation takes between 53 minutes and 40 hours.

\begin{table*}
\small\sf\centering
\caption{Time to compute 1 million steps}
\label{table:1M-steps}
 \begin{tabular}{ccccc}
 \toprule
 Resolution (m) & 8 cores & 32 cores & 128 cores & 512 cores \\ [0.5ex]
 \midrule
 8 & 43 mins & 13 mins & 5.9 mins & 6.1 mins \\
 4 & 2.7 hours & 44 min & 15 mins & 8.9 mins \\
 2 & 10 hours & 2.7 hours & 46 mins & 18 mins \\
 1 & 40 hours & 10 hours & 2.7 hours & 53 mins \\
 \bottomrule
\end{tabular}
\end{table*}

These results can be viewed in terms of weak and strong scaling efficiencies. The weak scaling efficiencies are shown in Table (\ref{table:weak-scaling}), and vary between 66\% and 100\%. The strong scaling efficiencies are shown in Table (\ref{table:strong-scaling}). At a given resolution, when e.g. quadrupling the number of cores, one would ideally reduce the computation time by 4. The strong scaling efficiencies indicate how close we get to the ideal. Unsurprisingly, they are largest in the region where the number of grid points per TPU is largest, so for example, in the 1 meter simulation, going from 8 cores to 32 cores, the run time decreases by a factor of 4 (99\% strong scaling). However, in the coarsest 8 meter simulation when going from 128 cores to 512 cores, the large communication overhead swamps the run time, so only an 11\% strong scaling efficiency is achieved.

\begin{table*}
\small\sf\centering
\caption{Weak scaling efficiencies}
\label{table:weak-scaling}
 \begin{tabular}{ccccc}
 \toprule
 Resolution (m) & 8 cores & 32 cores & 128 cores & 512 cores \\ [0.5ex]
 \midrule
  8 & \textcolor{gray}{43 mins} & \textcolor{gray}{13 mins} & \textcolor{gray}{5.9 mins} & \textcolor{gray}{6.1 mins} \\
 4 & \textcolor{gray}{2.7 hours} & \textcolor{gray}{44 min} 97\% & \textcolor{gray}{15 mins} 84\% & \textcolor{gray}{8.9 mins} 66\% \\
 2 & \textcolor{gray}{10 hours} & \textcolor{gray}{2.7 hours} 99\% & \textcolor{gray}{46 mins} 93\% & \textcolor{gray}{18 mins} 72\% \\
 1 & \textcolor{gray}{40 hours} & \textcolor{gray}{10 hours} 100\% & \textcolor{gray}{2.7 hours} 100\% & \textcolor{gray}{53 mins} 80\% \\
 \bottomrule
\end{tabular}
\end{table*}

\begin{table*}
\small\sf\centering
\caption{Strong scaling efficiencies}
\label{table:strong-scaling}
 \begin{tabular}{ccccc}
 \toprule
 Resolution (m) & 8 cores & 32 cores & 128 cores & 512 cores \\ [0.5ex]
 \midrule
  8 & \textcolor{gray}{43 mins} & \textcolor{gray}{13 mins} 83\% & \textcolor{gray}{5.9 mins} 46\% & \textcolor{gray}{6.1 mins} 11\% \\
 4 & \textcolor{gray}{2.7 hours} & \textcolor{gray}{44 min} 91\% & \textcolor{gray}{15 mins} 66\% & \textcolor{gray}{8.9 mins} 28\% \\
 2 & \textcolor{gray}{10 hours} & \textcolor{gray}{2.7 hours} 94\% & \textcolor{gray}{46 mins} 83\% & \textcolor{gray}{18 mins} 54\% \\
 1 & \textcolor{gray}{40 hours} & \textcolor{gray}{10 hours} 99\% & \textcolor{gray}{2.7 hours} 94\% & \textcolor{gray}{53 mins} 70\% \\
 \bottomrule
\end{tabular}
\end{table*}

\section{Broader Impacts}

In this paper, we have described the design and implementation of a library for scientific computation on TPUs, specifically as applied to the numerical direct integration of PDEs on regular Euclidean grids. While we have specialized the discussion in this paper to 2D riverine flood forecasting to highlight performance characteristics achievable in a concrete setting with clear impact on human safety, our approach directly generalizes to 3D models such as Navier-Stokes equations, turbulent downbursts, and ocean convection. In such settings with a greater number of interior boundaries and more complex boundary conditions and numerical integration schemes, the TPU ICI becomes even more critical in enabling good performance. Our libraries are readily applied to related problems and have been used for Fourier Transforms, Ising Model, and financial Monte Carlo, among others. The code used for this paper is open sourced at \url{https://github.com/google-research/google-research/blob/master/simulation_research/flood/}. We will open source more of our libraries to make scientific computing on TPU more accessible to users of Google Cloud Platform.

Hydraulic simulations such as those implemented in this paper are used by a wide range of relevant actors, for a diverse set of purposes. In this section we share a handful of common use-cases for such simulations. This list is not comprehensive.

Hydro-meteorological government agencies, as well as their parallels in international agencies and non-governmental organizations (and associated commercial hydrologic engineering companies which serve them) will often use such simulations for real-time flood forecasting. In these cases, these simulations may be run repeatedly at set intervals, initialized with either current conditions or forecasted data, providing services that range from nowcasting to forecasting weeks into the future. The outputs of these simulations will often be translated into human and economic impact (or risk), and then distributed to disaster management agencies, first responders and the general public\cite{nevo2019ml}.

These same simulations can also be used for water management and land management infrastructure planning. In this case, either engineering companies or relevant government agencies run hydraulic simulations to either plan and improve various aspects of water management they are building, or to test whether a proposed architecture would achieve its goals in hypothetical situations. As an example, before building a new embankment, one might run a hydraulic simulation of a 100-year return period discharge on an elevation map that includes the proposed embankment, to see if indeed flood harms are mitigated. This information would then be used either internally by the organization planning and building the infrastructure, or shared with relevant regulatory bodies that need to approve such plans.

In a similar fashion, these simulations are also useful for the real-time operation of existing water management infrastructure, such as dams. In this case, dam operators may run hydraulic simulations based on different discharges they could release themselves, to inform their decision-making. This allows them to identify the optimal release policy, balancing agricultural needs, electricity generation, and retainment of sufficient water for the dry season, while minimizing the risk of flooding or overflowing of the dam or reservoir. Once an optimal policy is identified, the dam operators can then execute that policy, and repeat the process as new information about upstream conditions becomes available.

Finally, such simulations are used by climate scientists and hydrologists to analyze the effects of different climate change scenarios on flooding. This is a critical step in translating climate models into concrete, localized costs, which are critical for climate-related decision-making. In this case, academics or environmental agency researchers will input the simulations with various hypothetical estimated basin conditions matching the different climate scenarios they are evaluating. They would then publish the results in various publications, or share them with policymakers to inform climate-related investments and mitigation or adaptation policies.

\section{Future Work}

TPUs are well suited for large matrix multiplications and for problems that involve those operations. There are ongoing efforts to increase the range of operations and improve the ease of access to more general computing on TPUs. Future work can tackle optimizing integration and flexibility between TPUs and CPUs and building better programming models for TPUs.

We hope that by running hydraulic simulations on TPUs, researchers will be able to explore directions that are currently hindered by computational costs. One example would be a data assimilation technique that aims to improve hydraulic models in data scarce regions \cite{shastry2019utilizing}.

With smaller, highly dynamic rivers, it is no longer reasonable to assume constant discharge rates and a steady state solution. Owing to the dynamic nature of the problem and the size of its parameter space, real time simulation is a particularly promising approach for providing timely information about inundation risk, as it can avoid prohibitively costly offline exhaustive parameter search and caching.

Future work can directly incorporate deep learning-based precipitation forecasts and other weather-derived inputs into flood simulations. Computing on TPUs, as well as distributed computing across multiple TPU cores, can also be applied to other scientific computing problems.

\section{Conclusion}

This paper demonstrates that TPUs can efficiently be employed to solve physical simulations. In particular, we modified our hydraulic inundation model, which was originally run on CPU, to run on TPU. We found speed increases on TPU of over 500$\times$ compared to CPU. We ran our model to simulate a flood in a section of the Arkansas River and found a discharge rate that resulted in good agreement with aerial imagery. We showed how the results scale with an increasing number of TPUs.

\subsection*{Acknowledgements}

The authors would like to thank Zvika Ben-Haim, Carla Bromberg, Niv Giladi, Zack Ontiveros, and Cliff Young.

\section*{Appendix - Code}

\subsection{Code Description}

This section describes the code\footnote{\url{https://github.com/google-research/google-research/tree/master/simulation_research/flood}} used to run physics-based simulations of floods using Tensor Processing Units (TPUs). The physics implemented are the shallow water equations in unidirectional form. This code is in a Jupyter notebook that can run on Google Colaboratory which can access Google Cloud TPUs.

\subsubsection{Inputs}

The input to the simulation are parameters defining the simulation area:
\begin{itemize}
\item $unpadded\_dem$ A rectangular elevation map of the simulation area, expressed as an 2D array.
\item $resolution$ The resolution of the elevation map.
\item $flux$ The discharge rate of the water entering the simulation area (units $m^3/s$).
\item $BoundarySide$ The side of the simulation area where water is entering or exiting.
\item $fraction\_start$ and $fraction\_end$ The (beginning and ending) fractions which define the region along a given side where water is entering or exiting.
\end{itemize}
In addition, there are parameters which describe how the simulation should be run, including:
\begin{itemize}
\item $dt$ The time step used for the simulation.
\item $start\_time\_secs$ and $num\_secs$ Start time in seconds and number of seconds to run the simulation.
\item $cx$ and $cy$ These describe how the computational grid is split up into subgrids. The $x$ axis is split $cx$ times, and the $y$ axis is split $cy$ times ($cx \times cy$ TPU cores are used in the computation).
\end{itemize}

\subsubsection*{Outputs}

The simulation calculates the height of the water and the flux in the $x$ and $y$ directions at each point in the 2D grid at each time step.

\subsubsection{Procedure}

All the inputs needed for the simulation shown in this paper (other than the DEM) are included with the code. The elevation maps are stored on Google Cloud Storage at 2, 4, and 8 meter resolutions.

\subsection{Code structure}

The method \texttt{run\_simulation} takes various inputs such as \texttt{resolution} and \texttt{flux}, as well as the DEM file name. It loads the DEM file from Google Cloud Storage. A mask for the portion of the map that corresponds to the river is created (the whole grid is used in this case), and a Manning matrix is computed from the river mask. The method then initializes parameters used to configure the simulation, and constructs the boundary conditions and some builders that are used to set up the computation. These are ultimately used as ingredients to construct a \texttt{TPUSimulationManager}, the overall unit that manages a simulation. In the following subsections, we explore some key abstractions.

\begin{figure*}
  \centering
  \includegraphics[scale=0.33]{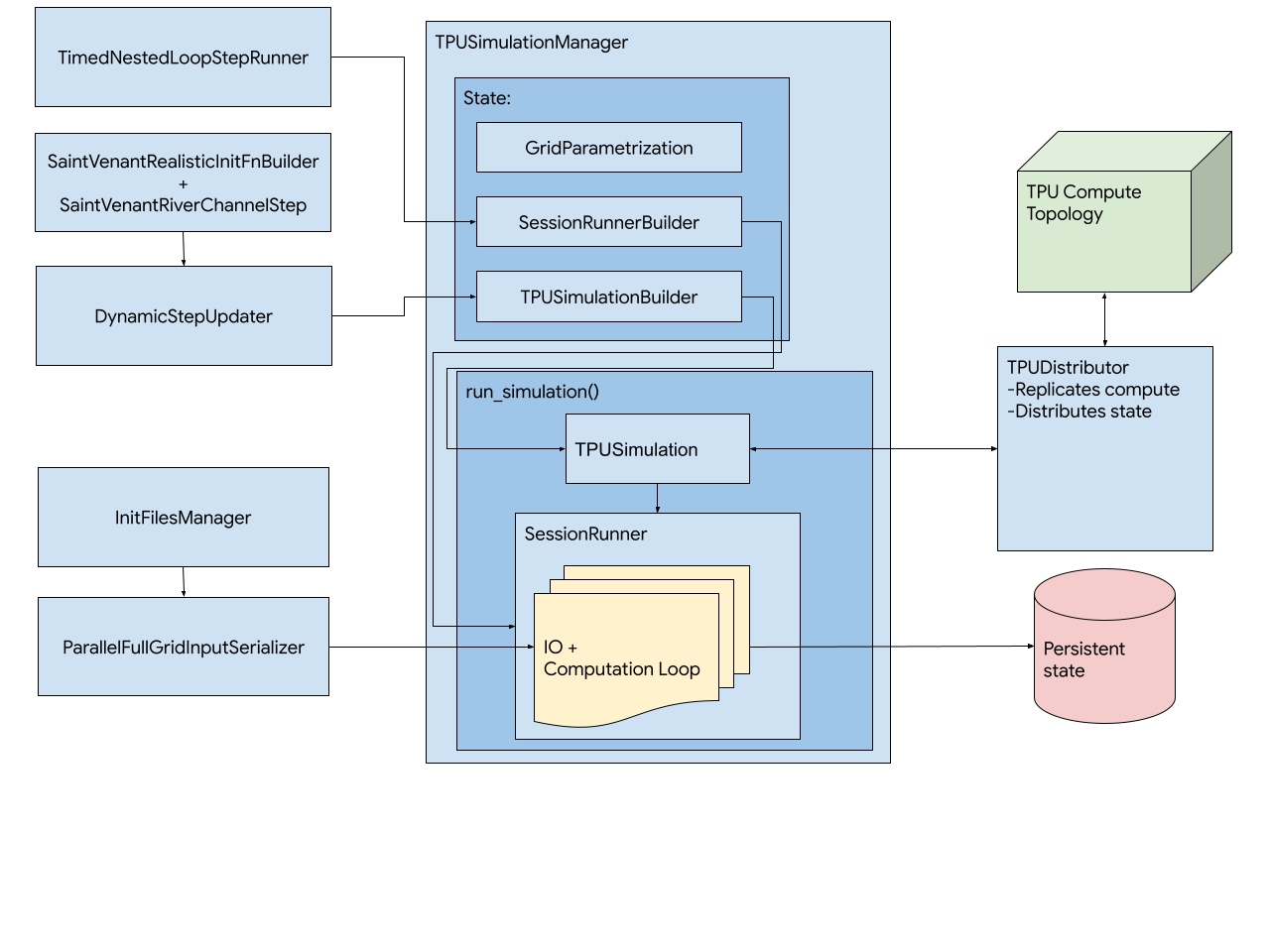}
  \caption{Class structure and assembly for building and running TPU simulations.}
  \label{fig:code_structure}
\end{figure*}


Briefly, \texttt{TPUSimulationManager} starts by creating a TPU-aware \texttt{tf.Session}. It then constructs a \texttt{TPUSimulation} using the session instance and a \texttt{TPUSimulation} builder (both described below), which is aware of the model update function and states to be used in the simulation. Next, it constructs the control flow units that will be responsible for running the steps of simulation, including reading in initial conditions and distributing these to the in-memory TPU representation, and running the temporal step iterations for the duration of the simulation. The control flow assembly is also responsible for outputting intermediate state results during the course of simulation, and reporting progress back to the master program.


\texttt{TPUSimulation} is a concrete class that manages the initialization of a simulation on TPU hardware. It makes use of \texttt{TPUDistributor}, an implementation detail of \texttt{TPUSimulation}, which manages the distribution of computation across a TPU grid. In particular, it manages replication of computation across TPUs, and the distribution and persistence of state across the grid. \texttt{TPUDistributor} constructs a control flow graph that uses \texttt{tf.tpu.replicate} to distribute the computation over each core for each core's mapped portion of the state space.


The \texttt{TPUSimulation} builder is a callable that accepts a \texttt{tf.Session} and returns a \texttt{TPUSimulation}. It is responsible for assembling the core computation, which it does by defining the initialization step (\texttt{SaintVenantRealisticInitFnBuilder}), the finite difference step (\texttt{SaintVenantRiverChannelStep}), and the timestepping control (\texttt{DynamicStepUpdater}).

\texttt{SaintVenantRealisticInitFnBuilder} is responsible for defining the constant and dynamic states in the simulation and reading DEM data from file. This unit is also responsible for initializing input and output flux using the boundary conditions instantiated in \texttt{InflowBoundaryCondition} and \texttt{OutflowBoundaryCondition}.

\texttt{SaintVenantRiverChannelStep} defines the Saint-Venant finite difference update. Within this unit, at each time step, we first compute boundary conditions. We then perform halo exchange on the flux states to ensure that the outer values for the portion of the state space mapped to each core incorporate the adjacent values from the neighboring cores with the boundary conditions applied, so that the finite differences at the outer values are computed correctly. We then perform the finite difference update.
The finite difference update can be implemented in a batched fashion using
TensorFlow's \emph{slice} operation. The \emph{slice} operation extracts a sub-array from a Tensor given the starting index and the size of the desired slice along an axis. An alternative to using \emph{slice} operations would be TensorFlow's \emph{convolution} operation. The \emph{convolution} operation moves a sliding window of coefficients called the \emph{kernel}.  At each position of the input array, the computed output is the dot product of the kernel and overlapping region of the input array. Convolutions are attractive on TPUs since they are highly optimized. However, we found that for low order finite difference updates, slice operations are faster than convolutions.

Finally, we do another halo exchange to propagate the new \texttt{h} values. Note that for the very first call, the halos are correct by initialization. In subsequent calls, the \texttt{h} halos are correct by virtue of the deferred \texttt{h} halo exchange that happens immediately after each finite difference update.

Halo exchange is performed by having each TPU core select a border row from its subgrid. A \texttt{collective\_permute} operation is called which both sends the row to and receives the corresponding row from the neighboring TPU. This is done for each border, in turn. In case the border is a global border, border value is either directly computed (if the border includes an inflow or outflow region) or is unchanged.

\texttt{DynamicStepUpdater} defines a \texttt{tf.while\_loop} control flow that steps the simulation forward for a given number of seconds with a given base \texttt{dt}, adjusting the final timestep so as not to exceed the desired number of seconds for that epoch of simulation.

The remaining orchestration is straightforward, and consists of setting up the \texttt{TimedNestedLoopStepRunner}, which runs the simulation for the desired length of time and outputs intermediate progress. We refer the reader to \ref{fig:code_structure} for a visual representation and summary of how the abstractions we have described fit together.

\subsection{Running the notebook}

To run the notebook on Google Cloud TPUs, first download the elevation map. You will also need a Google Storage Bucket to write output files. Then follow these steps:
\begin{enumerate}
    \item Navigate to \url{https://colab.research.google.com/}
    \item Go to File $>$ Open notebook. Navigate to open a notebook on GitHub. Use the following GitHub URL: \url{https://github.com/google-research/google-research/blob/master/simulation_research/flood/flood_simulation_tpu.ipynb} and open that notebook.
    \item The Jupyter notebook should now be loaded. Click the 'Connect' button on the top right to connect to a runtime in the cloud.
    \item Run each cell consecutively. At the end, a visualization will be generated looking down on the simulation area showing the areas that are flooded.
\end{enumerate}

\bibliographystyle{TRR}
\bibliography{bibliography}

\end{document}